\documentclass[prl,nofootinbib,twocolumn]{revtex4}
\usepackage{graphics}
\usepackage{bm}

\begin{document}

\title{511~keV Photons from Superconducting Cosmic Strings}

\author{Francesc Ferrer and Tanmay Vachaspati
}
\affiliation{
CERCA, Department of Physics, Case Western Reserve University,
10900 Euclid Avenue, Cleveland, OH 44106-7079, USA.}

\begin{abstract}
\noindent We show that a tangle of light superconducting strings 
in the Milky Way could be the source of the observed 511~keV emission
from electron-positron annihilation in the Galactic bulge. The
scenario predicts a flux that is in agreement with observations
if the strings are at the $\sim 1$~TeV scale making the particle 
physics within reach of planned accelerator experiments. The emission 
is directly proportional to the galactic magnetic field and future 
observations should be able to differentiate the superconducting
string scenario from other proposals.
\end{abstract}

\maketitle

The detection of a bright 511~keV line by the SPI spectrometer on the 
International Gamma-Ray Astrophysics Laboratory (INTEGRAL) satellite, 
has established the presence of a diffuse source of positrons in the 
Galactic Center (GC)~\cite{SPI}. The observed photon flux of 
\begin{equation}
9.9^{+4.7}_{-2.1} \times 10^{-4}~{\rm cm}^{-2} {\rm s}^{-1}
\label{f0}
\end{equation} 
with a linewidth of about 3~keV is in good agreement with previous 
measurements~\cite{osse}. The observations suggest a spherically 
symmetric distribution. Assuming a Gaussian spatial distribution 
for the flux, a full width at half maximum of $9^\circ$ is indicated.

The origin of these Galactic positrons remains a mystery. Several
scenarios involving astrophysical sources have been proposed, including 
neutron stars or black holes, massive stars, supernovae, hypernovae, 
gamma-ray bursts or cosmic rays~\cite{astrof}. However, the fraction of 
positrons produced in such processes is uncertain, and it is unclear that 
the positrons could fill the whole bulge. 

Alternatively, mechanisms associated with the Dark Matter~(DM) at the GC 
have been put forward. If DM is constituted by a light (1-100~MeV) scalar, 
its annihilations could account for the observed signal~\cite{Boehm:2003bt}. 
Another possibility is that DM could be in the form of nonhadronic color 
superconducting droplets. Positrons on the surface of the droplet could 
annihilate with ambient electrons producing a gamma-ray 
line~\cite{Oaknin:2004mn}. 

In this Letter, we show that a network of light superconducting strings 
\cite{Witten:1984eb} occurring in particle physics just beyond
the standard model could also be a source of positrons. More detailed
observations of the positron distribution could be used to distinguish
this source from the other possibilities. Moreover, this resolution of 
the positron observation works with strings at an energy scale of 
$\sim 1$~TeV, since the flux from heavier strings is lower. Therefore, 
experiments at the Large Hadron Collider in the near future will also 
probe particle physics relevant for this explanation of the positrons.

We assume that a tangle of light superconducting strings exists 
in the Milky Way. 
The strings are frozen in the Milky Way 
plasma as long as the radius of curvature is larger than a certain 
critical length scale. If the curvature radius is smaller, the 
string tension wins over the plasma forces and the string
moves with respect to the magnetized plasma.
During the string motion, the loop will cut across the Milky
Way magnetic field, generating current as given by Faraday's
law of induction. The current is composed of zero modes of charged 
particles, including positrons, propagating
along the string. Once the energy of a positron on the 
string exceeds the rest energy of 511~keV, it becomes possible 
for it to leave the string primarily due to scattering by 
counter-propagating particles. Even though the scattering rates 
are model dependent, general arguments discussed below
\cite{Barr:1987ij}
indicate that the scattering of positrons with counter-propagating 
light quarks will be very efficient at ejecting the positrons at the 
threshold of 511~keV. The ejected positrons will annihilate with the 
ambient electrons, thus emitting 511~keV gamma rays.

The string dynamics is determined by comparing the force due to
string tension to the plasma drag force. The force per unit length 
due to string tension $\mu$ is
\begin{equation}
F_s \sim \frac{\mu}{R},
\end{equation}
where $R$ is the string radius of curvature. The drag force is more 
complicated, needing an analysis of the plasma flow around a current 
carrying string \cite{Chudnovsky:1986hc,VilShe94}. The string
effectively behaves like a cylindrical body around which the
plasma flows. The radius of the cylinder, in natural units, is 
$r_s = {J}/({v_{\rm rel} \sqrt{\rho}})$, $J$ being the current 
on the string, $v_{\rm rel}$ the velocity 
of the string relative to the plasma, and $\rho$ the plasma 
density. The drag force on such a cylinder is
\begin{equation}
F_d \sim \rho v_{\rm rel}^2 r_s \sim \sqrt{\rho} v_{\rm rel} J.
\end{equation}
The maximum value of the damping force occurs when $v_{\rm rel} =1$. 
Therefore, $F_{{\rm d},{\rm max}} \sim \sqrt{\rho} J$.
If $F_s \gg F_{{\rm d},{\rm max}}$, the string moves under its own tension 
and the damping force is insufficient to prevent the string from
accelerating to relativistic velocities. String loops will then emit 
electromagnetic radiation and eventually dissipate. If, however, 
$F_s < F_{{\rm d},{\rm max}}$, the string will accelerate until the damping
force grows so as to cancel the tension. Then the string moves at
a terminal velocity relative to the plasma found by equating $F_s$ 
and $F_d$:
\begin{equation}
v_{\rm term} \sim \frac{\mu}{\sqrt{\rho} JR}.
\label{vterm}
\end{equation}
Setting $v_{\rm term} =1$, we find the critical radius of curvature
$R_c$ when the strings move relativistically:
\begin{equation}
R_c \sim \frac{\mu}{\sqrt{\rho} J}.
\end{equation}
To summarize, strings with tension $\mu$, current $J$, and curvature 
radius $R> R_c$, move at $v_{\rm term}$ with respect to a plasma 
of density $\rho$.  If $R < R_c$, the string moves at relativistic 
speeds.

In a turbulent plasma, such as in our Milky Way, there is another 
length scale of interest, called $R_*$ ($R_* > R_c$), even when the 
string motion is overdamped. For $R > R_*$, the terminal speed of
the strings is small compared to the turbulence speed of the plasma
and the strings are carried along with the plasma. As the strings
follow the plasma flow, they get more entangled due to turbulent
eddies, and the strings
get more curved until the curvature radius drops below $R_*$. Then
the string velocity is large compared to the plasma velocity, and, 
hence, the strings break away from the turbulent flow. Therefore,
$R_*$ is the smallest scale at which the string network follows
the plasma flow. For $R_* > R > R_c$, the string motion is over-damped
but independent of the turbulent flow. Hence, string curvature on 
these scales is not generated by the turbulence, and we can estimate
the length density of strings in the plasma as $\rho_l \sim {1}/{R_*^2}$.
To find $R_*$, denote the turbulent velocity at the scale $R_*$ by 
$v_*$ and set $v_{\rm term} \sim v_*$, to get \cite{Chudnovsky:1988cv}
\begin{equation}
v_* R_* \sim \frac{\mu}{\sqrt{\rho} J} 
      = \sqrt{\frac{\mu}{\rho}} \frac{1}{e \kappa},
\label{v*R*}
\end{equation}
where, for convenience, the dimensionless parameter $\kappa$ has been 
introduced via $J \equiv \kappa e \sqrt{\mu}$. 

To determine $v_*$, we note that, on scales less than 
$l \sim 100 {\rm pc} = 3 \times 10^{20} {\rm cm}$ \cite{ZelRuzSok83},
the interstellar plasma has velocity spectrum given by 
magnetohydrodynamic turbulence as $v_R \sim v_l ( R/l )^{1/4}$, 
where $v_l \sim 10^6$ cm/s. Inserting this expression in Eq.~(\ref{v*R*})
and solving for $R_*$ gives
\begin{equation}
R_* \sim l 
   \left ( \sqrt{\frac{\mu}{\rho}} \frac{1}{e \kappa v_l l} \right )^{4/5}
~ , \ \ 
v_* \sim v_l 
    \left ( \sqrt{\frac{\mu}{\rho}} \frac{1}{e \kappa v_l l} \right )^{1/5}.
\label{Rstar}
\end{equation}
Note that in these estimates we have assumed $R_* < l$; otherwise,
the turbulent scaling law will be different. 



The current on the string is generated by the motion of the string across the 
galactic magnetic field as described by Faraday's law of induction and is
limited by the microscopic interactions of the charge carriers on the string.
A given charge carrier can, in principle, leave the string once it has enough 
energy. The escape is triggered by several factors,
of which the most important is scattering off counter-propagating
zero modes (bosons or fermions). The current in any particular species
saturates once the scattering rate equals the growth rate due to Faraday 
induction. The rate of current growth for particle species (denoted $X$) 
with charge $e$ is given by ${\dot J}_X \sim e^2vB$. The time scale for current
growth, assuming $v \sim c$, is
\begin{equation}
\tau_{\rm growth}^X \sim \frac{J_X}{e^2 B}.
\label{taugrowth}
\end{equation}

The current decays because the charge carriers on the string can interact 
and leave the string. 
The positron current is limited mainly by scattering with
counter-propagating quarks ($u$ quarks for electroweak strings).
Since the lightest quarks have $\sim 1$~MeV mass, which is similar
to the positron mass $m_e$, the decay time when the positron current is 
order $e m_e$ is $\tau_e = 1/n \sigma$, 
where $n \sim (J_e/e) m_e^2$ is the number density of positrons and  
$\sigma \sim \alpha^2 /m_e^2$ is the electromagnetic scattering 
cross section, $\alpha$ being the fine structure constant. Therefore,
$\tau_e \sim e \alpha^{-2} J_e^{-1}$.
Equating $\tau_e$ and the growth rate given by $\tau_{\rm growth}^e$,
we get $J_e \sim \sqrt{B} \sim 10^{-4}$~eV,
with $B \sim 10^{-6}$~G and $\kappa \sim 1$.
Since this is much less than $m_e = 511$ keV and positrons cannot escape 
the string until they reach 511 keV energy, the threshold for their escape 
is also 511 keV. Here we should note that the quark that scatters off
the positron cannot completely escape from the string unless it gains 
$\sim 100$~MeV energy and hadronizes. However, at energies less than 
100~MeV, the quark can still get kicked out from its zero mode state 
on the string to a quark that is not in a zero mode. Such a quark will 
be confined to a $(\Lambda_{\rm QCD})^{-1} \sim (100~{\rm MeV})^{-1}$
shell around the string. Pion emission from the string cannot 
deplete the baryonic current on the string. 
Only at $\sim 1$~GeV energies can 
antiprotons be emitted, leading to another possible signature of 
galactic superconducting strings \cite{Starkman:1996hi}.

As a piece of string of length $R_*$ cuts through a magnetic field
$B$, it will produce electrons or positrons, with equal likelihood,
at the rate $d N / dt \sim e v_* B R_*$.
In a volume $V=4\pi L^3/3$, there are $\sim L^3/R_*^3$ such pieces 
of string, and, hence, the rate of particle production in the entire volume 
is
\begin{equation}
\frac{d N_{\rm V}}{dt} \sim e v_* B \frac{L^3}{R_*^2}.
\label{dNVdt}
\end{equation}
The current in the positrons will grow at first but then saturate
at 511 keV. 
After that, further motion of the string across the galactic magnetic
field will generate positrons that leave the string. So $N_V$ is also the 
number of positrons being produced in the volume $V$ which we denote
by $N_+$. Inserting Eq.~(\ref{Rstar}) in (\ref{dNVdt})
we get
\begin{equation}
\frac{d N_+}{dt} \sim e^{12/5} B \kappa^{7/5} \frac{L^3}{l^3} 
                      \left ( \frac{\rho}{\mu} \right )^{7/10}
                       ( v_l l )^{12/5}.
\label{dN+dt}
\end{equation}
We are interested in a region of radius 1 kpc around
the galactic center and so $L^3 \sim V_1 (1 {\rm kpc})^3$, where
$V_1$ is a dimensionless parameter. The plasma 
density in the galactic center is higher than the average
Milky Way density. For an estimate, we adopt an isothermal sphere 
scaling normalized to $\rho_8 \sim 10^{-25} {\rm gm/cm^3}$ 
at 8 kpc from the center, thus giving 
$\rho \sim 6 \cdot 10^{-24} \rho_{\rm gc}\, {\rm gm/cm^3}$, where we 
have introduced the parameter $\rho_{\rm gc}$.
The magnetic field in the galactic center is not known very 
precisely but is estimated to be $\sim 10^{-3}$ G~\cite{Chandran:2000rg}. 
Therefore, we 
will set $B = B_3 10^{-3}$ G. (Note the conversion of magnetic  
field strength in Gauss to particle physics units:  
$1 {\rm G} = 1.95\times 10^{-20}~{\rm GeV^2}$.) The parameters
$v_l$ and $l$ will be different in the galactic center but
these are not known. We will write $v_l=10^6~v_{l,6}$ cm/s 
and $l=100 ~ l_{100}$ pc.  The string tension 
will be taken to be $\mu = \mu_1 (1~{\rm TeV})^2$, where $\mu_1$ is a
parameter, and $e^2 \sim 0.1$, and then Eq.~(\ref{dN+dt}) gives us
\begin{equation}
\frac{d N_+}{dt} \sim 
     10^{42} B_3 \kappa^{7/5} V_1 \mu_1^{-7/10} \rho_{\rm gc}^{7/10} 
        v_{l,6}^{12/5} l_{100}^{-3/5}\;{\rm s}^{-1}.
\label{positronrate}
\end{equation}
Although the astrophysical parameters describing the galactic
center are not known very accurately,
assuming equipartition of plasma kinetic energy ($\sim \rho v_l^2$) and 
magnetic energy ($\sim B^2/8\pi$), with $l \sim \rho^{-1/3}$, we find that
$v_{l,6} \sim 100$ and $l_{100} \sim 0.1$, which boosts the estimate in 
Eq.~(\ref{positronrate}) by $10^5$, yielding 
\begin{equation}
\frac{d N_+}{dt} \lesssim 10^{47}~ {\rm s}^{-1}.
\label{numericalestimate}
\end{equation} 

From Eq.~(\ref{Rstar}), the string velocity with respect to the plasma 
is $v_*\sim 10^{-5} c$. Hence, the positrons will leave the string with a small
Lorentz factor, $\gamma \sim 1$, and they will gradually slow down
by Coulomb collisions in the interstellar medium (ISM).
The energy loss rate is 
approximately~\cite{Longair} $dE/dt \sim 2 \times 10^{-9} 
(N_{H}/10^5 {\rm m^{-3}})
(\log \gamma+6.6) {\rm eV/s}$, where $N_{H}$ is the number density of
target atoms. This rate yields a stopping 
distance of $10^{24} {\rm cm}$. Considering a simple random walk, with 
the Larmor radius on the order of $10^1 {\rm cm}$, the positrons 
travel a distance $\sim 10^{12} {\rm cm}$, so they are easily
 confined to the galactic center. Note that this distance is less
than the typical string separation estimated from Eq.~(\ref{Rstar}),
$R_*\sim 5\cdot 10^{15} {\rm cm}$, so the positrons will 
have linelike features on angular scales $\sim 10^{-7}$, which 
is too small to be resolved with the $2^\circ$ angular precision of INTEGRAL.
Once produced, the positrons will undergo different processes in the ISM.
Pair annihilation with an ambient electron and positronium (Ps) formation and
decay via para-Ps will occur both in-flight and after thermalization. The
resulting spectrum depends on the specific details of the ISM, but detailed
analysis shows that a narrow 511~keV line generically results, 
$\Delta E \sim$3~keV, in agreement with observations~\cite{guessoum}. Hence,
the spectral shape does not depend on the details of the strings, and in
this respect the model cannot be distinguished from alternative mechanisms
such as light DM or astrophysical sources.

Three-quarters of ortho-Ps
annihilate in a 3 photon continuum final state.
As a result, each positron will contribute $(2-3f_{Ps}/2)^{-1}$ 511~keV 
photons, where the Ps fraction has been measured to be $f_{Ps}=0.93\pm0.04$
for the galactic center~\cite{Kinzer:2001ba}. 
Multiplying the observed gamma-ray flux given in 
Eq.~(\ref{f0}) by the area of the sphere at our location (8 kpc)
around the galactic center, we get the actual 
positron production rate in the galactic center:
\begin{equation}
\frac{d N_+^{obs}}{dt} \sim 
 \frac{1}{2-3f_{Ps}/2} 10^{-3} 
   \frac{4 \pi ( 8~{\rm kpc} )^2}{{\rm cm}^{2} ~{\rm s}} 
  \sim 1.2 \times 10^{43}\;{\rm s}^{-1}.
\label{observedrate}
\end{equation} 
Comparing Eqs.~(\ref{numericalestimate}) and (\ref{observedrate}),
we conclude that light superconducting strings are possible sources
of positrons that lead to the flux of 511~keV gamma rays observed by 
the INTEGRAL collaboration.

We see from Eq.~(\ref{positronrate}) 
that a unique prediction of our scenario is that the gamma-ray 
flux is proportional to the magnetic field strength in the
Milky Way, with a milder dependence on the plasma density. In the disk, 
the magnetic field intensity decreases by 
$B_3 \sim 10^{-3}$. At the same time, looking towards
the disk, the volume of the Milky Way that contributes 
positrons is larger. Taking these factors into account, for 
a disk with thickness 1~kpc and a radial extent of 30~kpc,
we estimate a photon flux 
$\sim 10^{-6}~{\rm cm}^{-2}{\rm s}^{-1}$ in a $16^\circ$ field
of view as in SPI. In a direction perpendicular to the disk, 
the volume contributing to the flux will be smaller, 
adding to the suppression of the magnetic field to yield 
a flux of $\sim 10^{-7}~{\rm cm}^{-2}{\rm s}^{-1}$.
Thus far, there have been no reliable detections outside of the 
central region of our Galaxy, with SPI placing an upper bound of 
$1.2 \times 10^{-4}~{\rm cm}^{-2}{\rm s}^{-1}$ on the 
flux~\cite{Teegarden:2004ct}. 
The target sensitivity of the SPI instrument, once 
sufficient exposure becomes available, is $2\times 10^{-5} ~ 
{\rm cm}^{-2}{\rm s}^{-1}$ at 511~keV ~\cite{spi}, somewhat 
above what is needed to map the emission from the 
disk in our scenario. 

That the flux should follow the magnetic field is in marked 
contrast with the MeV DM hypothesis. There the flux follows 
$\rho_{\rm DM}^2$, and a signal from nearby DM dominated regions, 
e.g., the Sagittarius dSph galaxy, is expected~\cite{Hooper:2003sh}. 
No magnetic fields have been measured in Sagittarius, although low 
surface brightness galaxies, which are somewhat similar, show 
$\mu G$ scale magnetic fields~\cite{klein}. Therefore, if 
superconducting strings source the observed 511~keV, the estimate
would be weaker by $10^{-3}$ than that of Eq.~(\ref{positronrate}) due 
to the weaker magnetic field, and another factor $\sim 10^{-1}$ due
to the source being 3 times further away. The volume of
Sagittarius dSph is comparable to the Milky Way bulge, leading 
to, at most, a flux of $\sim 10^{-7}~{\rm cm}^{-2} {\rm s}^{-1}$ 
in the direction of Sagittarius, some three orders of magnitude 
fainter than the MeV 
DM model prediction. After two Galactic Center Deep Exposures, 
INTEGRAL has not detected emission from Sagittarius, although the 
effective observation time is not yet sufficient to reach the 
sensitivity of the MeV DM predicted fluxes~\cite{Cordier:2004hf}. 
For the color superconducting DM scenario, the flux follows 
$\rho_{\rm visible}\rho_{\rm DM}$~\cite{Oaknin:2004mn}, 
which is also different from our proposed scenario.


Present-day galactic magnetic fields could have been produced by 
amplification of a tiny primordial field by a galactic dynamo. 
In that case, currents could build up in the network of superconducting 
strings at earlier epochs. Since there is no turbulence at 
last scattering, a tangle of strings will not be formed; 
there is only $\sim 1$ string per horizon 
which would simply be
dragged by the plasma.
During recombination,
the plasma density drops, and the damping force on strings is 
vastly reduced. String tension
causes strings to move at relativistic speeds cutting across magnetic fields 
and generating a flux of positrons, which will once 
again annihilate with ambient electrons
to produce 511 keV gamma rays. By the present epoch, the gamma
rays would have redshifted to 511 eV lines. 
One string per horizon gives a positron flux
of $\sim 10^{-17} ~{\rm cm^{-2}} {\rm s^{-1}} {\rm sr^{-1}}$, much
smaller than the upper limit on the diffuse extragalactic component by the 
Chandra experiment~\cite{Markevitch:2002dt} of 
$10^{-3} {\rm s^{-1} cm^{-2} sr^{-1}}$ at 511 eV, assuming a
redshifted width of 3~eV.

Recently, the High-Energy Antimatter Telescope (HEAT)
balloon experiment has confirmed an excess of high-energy
positrons in cosmic rays at energies around 10~GeV~\cite{HEAT}. 
Direct production of 10~GeV positrons from superconducting strings
does not seem likely, because it is hard for the positron 
current to build up beyond 1~MeV in the presence of counterpropagating 
zero modes. However, superconducting
strings at the TeV scale could still produce heavy charged 
fermions, if such zero modes exist, which could then decay to
give positrons in the 10~GeV energy range. The production rate of 
heavy fermions can be computed from Eq.~(\ref{positronrate}) to be
$\sim 10^{-23} ~{\rm cm^{-3}} {\rm s^{-1}}$. Although the actual 
positron yield depends on the branching ratio in the particle 
physics model and the details of the diffusion process, 
comparing this value to the 
typical values of $\sim 10^{-27} ~{\rm cm^{-3}} {\rm s^{-1}}$ 
for positrons from annihilations of dark matter particles at the 
electroweak scale~\cite{Kamionkowski:1990ty} shows that this 
possibility deserves further investigation.

We conclude that light superconducting strings could produce
enough positrons in the galactic center to explain the flux
of 511 keV gamma rays observed by the INTEGRAL collaboration.
The scenario can be differentiated from other proposals by
higher resolution observations and by observations in directions
away from the galactic center. Light superconducting strings
might also produce stellar and cosmological signatures. Since
the strings may be at the 1~TeV energy scale, the involved 
particle physics is within the energy range of planned accelerator 
experiments.

\begin{acknowledgments} 
We are grateful to Ed Witten for motivating this work with his 
question as to whether superconducting strings may provide an 
explanation for the 10~GeV positrons observed by the HEAT 
collaboration.
We are grateful to Corbin Covault, Wayne Hu, Mathieu Langer, 
Jim Peebles, Glenn Starkman and Alex Vilenkin for comments and 
discussions. This work was supported by the U.S. Department of 
Energy and NASA at Case.
\end{acknowledgments}

\end{document}